\documentclass[final]{svjour3}
\usepackage{graphicx}
\usepackage{rotating}
\usepackage{amssymb}
\usepackage{mathptmx}
\usepackage{amsmath}
\usepackage{subfigure}

\usepackage[numbers]{natbib}
\makeatletter

\bibpunct{}{}{,}{s}{}{,}

\begin{document}

\newcommand{\hdblarrow}{H\makebox[0.9ex][l]{$\downdownarrows$}-}
\title{Electron-phonon coupling in Ti/TiN MKIDs multilayer microresonator}



\author{M.~Faverzani$^{1,2}$ \and P.K.~Day$^{3}$ \and E.~Ferri$^{2}$ \and A.~Giachero$^{1,2}$ \and B.~Margesin$^{4,5}$ \and R.~Mezzena$^{5,6}$ \and A.~Nucciotti$^{1,2}$ \and A.~Puiu$^{1,2}$}

\institute{1: Dipartimento di Fisica, Universit\`{a} di Milano-Bicocca, Milano, Italy\\
2: Istituto Nazionale di Fisica Nucleare, Sezione di Milano-Bicocca, Milano, Italy\\
3: Jet Propulsion Laboratory, Pasadena, CA, U.S.A.\\
4: Fondazione Bruno Kessler, Trento, Italy\\
5: Istituto Nazionale di Fisica Nucleare, Trento Institute for Fundamental Physics and Applications (TIPFA), Trento, Italy\\
6: Dipartimento di Fisica, Universit\`{a} di Trento, Povo (TN), Italy\\
\email{marco.faverzani@mib.infn.it}}

\maketitle

\begin{abstract}

Over the last few years there has been a growing interest toward the use of superconducting microwave microresonators operated in quasi-thermal equilibrium mode, especially applied to single particle detection. Indeed, previous devices designed and tested by our group with X-ray sources in the keV range evidenced
that several issues arise from the attempt of detection through athermal quasiparticles produced within direct strikes of X-rays in the superconductor material of the resonator.
\\In order to prevent issues related to quasiparticles self-recombination and to avoid exchange of athermal phonons with the substrate, our group focused on the development of thermal superconducting microresonators. In this configuration resonators composed of multilayer films of Ti/TiN sense the temperature of an absorbing material. To maximize the thermal response, low critical temperature films are preferable. By lowering the critical temperature, though, the maximum probing power bearable by the resonators decrease abruptly because of the weakening of the electron-phonon coupling. A proper compromise has to be found in order to avoid signal to noise ratio degradation. In this contribution we report the latest measurement of the electron-phonon coupling.


\end{abstract}

\section{Introduction}

The superconducting microwave microresonator\cite{PDAYNATURE} constitute a promising technology for single particle detection both in terms of single detector performance and, especially, in multiplexability. Despite the growing applications in bolometric experiments, the single particle detection mode is not yet competitive respect to other more mature detection technologies.\\
In the past our group attempted to adapt the superconducting microwave microresonators to the single particle detection in the keV region through direct strike of X-rays in the sensitive part of the resonator; this work, however, evidenced several issues which arise when a relatively large density of energy is created as a consequence of an energetic event absorbed directly by the superconductor which constitutes the resonator\cite{MIATESI}.
\\To overcome these complications, an alternative detection mode can be considered. Given the equivalence of the effects on a superconductor due to a temperature variation and external pair breaking\cite{GAOEQUIVAL}, superconducting microwave microresonators can be designed to sense the variation of temperature of an absorber thermally linked to the resonator. This technique has been already demonstrated to work with TiN$_x$ resonators coupled with a Ta absorber achieving an energy resolution of 75 eV on the 5.9 keV K$_\alpha$ line of Mn\cite{ULBRICHT}.
\\Our purpose is to design quasi-thermal equilibrium detectors made of Ti/TiN multilayer, which can provide a valuable technique to produce films with adjustable critical temperature in the (0.5 $\div$ 4.6) K range\cite{ANDREA_TIN}. Such resonator will be thermally coupled with an absorber where energetic events release all their energy. In order to obtain a sensitive variation of temperature of the absorber, this must be kept at a temperature as low as possible, namely below $\sim$ 100 mK. On the other hand, resonators present their steepest responsivity to temperature variations at temperatures above $\sim$ $T_\text{c}/4$, so very low critical temperatures are preferable.

\section{Resonators geometry}\label{sec:sec2}

\begin{figure}[htbp]
\begin{center}
\includegraphics[width=0.5\linewidth, keepaspectratio]{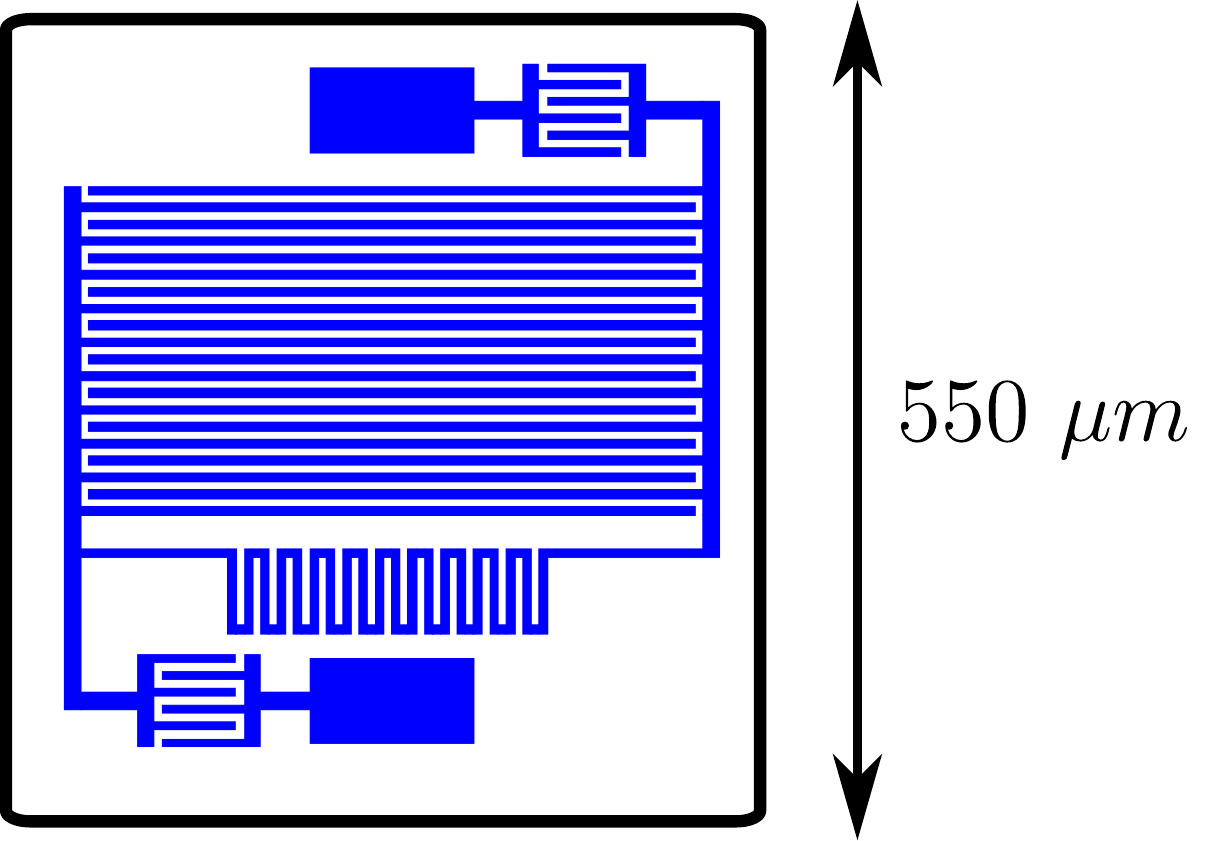}
\caption{Lumped element resonator used for thermal quasi-equilibrium. The original 600 $\mu$m thick bulk silicon substrate was reduced at about 100 $\mu$m. The reduced thickness and the small linear dimensions are intended to contain as much as possible the heat capacity.}
\label{fig:fig1}
\end{center}
\end{figure}

We produced resonators in the lumped element form (Fig.~\ref{fig:fig1}) composed of Ti/TiN multilayer with a critical temperature of 640 mK deposited on a bulk silicon substrate 600 $\mu$m thick. This film, composed of 12 layers of Ti/TiN 10 and 7 nm thick, respectively, features a sheet inductance of 16 pH/$\square$ with a total thickness of 204 nm. This relatively low critical temperature combined with this large kinetic inductance make this film a good candidate for thermal quasi-equilibrium detection.

\begin{figure}[htbp]
\begin{center}
\subfigure[]{
\includegraphics[height=0.425\linewidth, keepaspectratio]{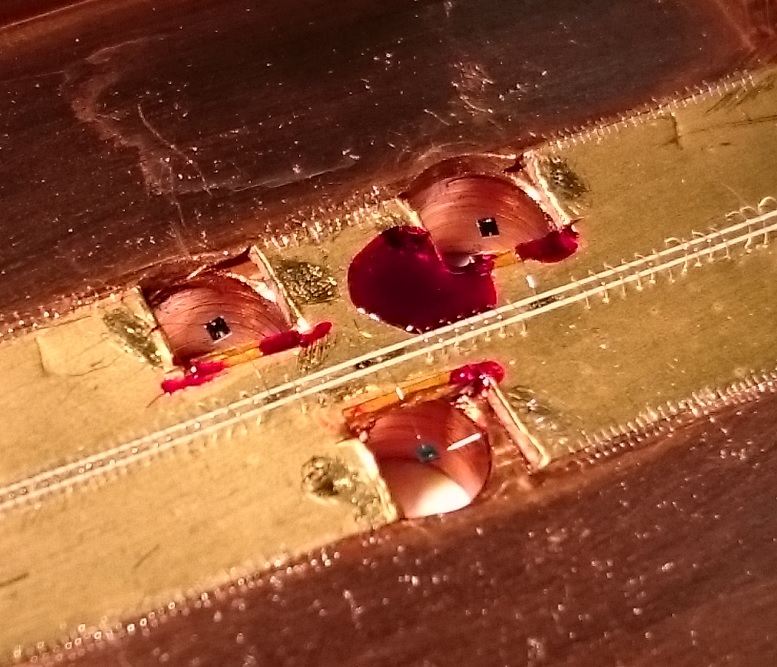}\label{fig:fig2a}
}
\subfigure[]{
\includegraphics[height=0.425\linewidth, keepaspectratio]{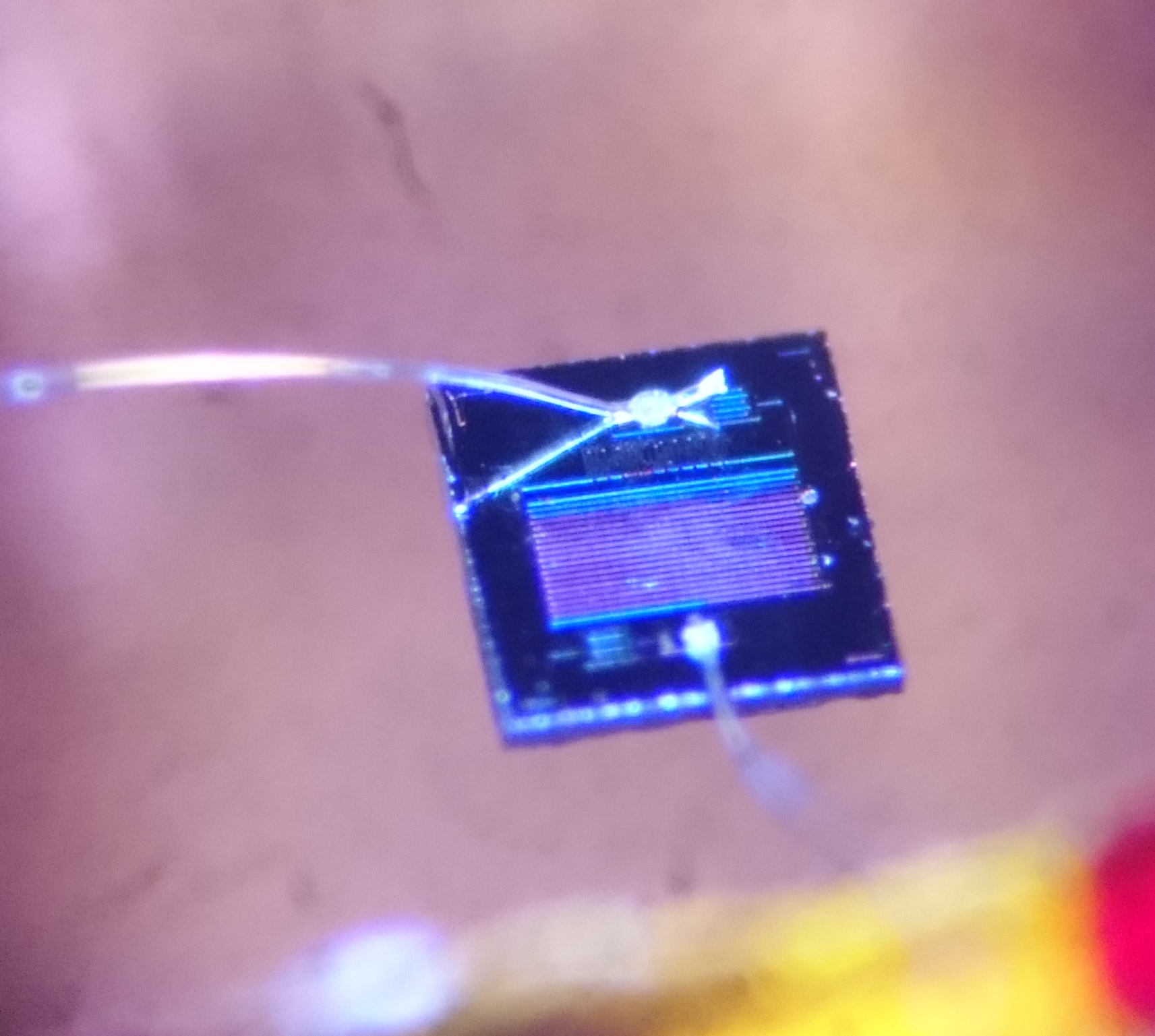}\label{fig:fig2b}
}
\caption{{\it Left} The gold CPW used as feedline for three suspended chips. {\it Right} Zoom in of one single chip suspended by means of two 17 $\mu$m Al/Si 1\% bonding wires: these provide a finite thermal link, mechanical support and electrical contacts.}
\label{fig:fig2}
\end{center}
\end{figure}

As thermal reference to the the bath, mechanical support and electrical contacts two 17 $\mu$m Al/Si 1\% bonding wires were used, so that the chip was kept suspended very close to a gold CPW used as feedline. Of the two bonding wires, one ran from the central part of the CPW to one pad of the chip while the second one ran from the ground plane to the other pad (Fig.~\ref{fig:fig2}). In this configuration a $^{55}$Fe X-ray test source was faced to the bottom side of the chips so that the silicon substrate would act as X-ray absorber. A X-ray strike would cause a temperature rise of the silicon which would be detected by the resonator. In order to maximize the thermal response the substrate thickness was reduced to about 100 $\mu$m.

\section{Measurements}

The first measurement regarded the steady-state characterization of the resonators. By comparing the measured resonant frequency to the simulated one with Sonnet it was possible to establish that the sheet inductance of the film was 16 pH/$\square$. The gap parameter was determined by fitting the internal quality factor as a function of the temperature (procedure described in \cite{MIATESI}) and it was found to be 0.09 meV. The measured internal quality factors were around $6\cdot 10^{3}$, almost two orders of magnitude smaller that what we have found with Ti/TiN multilayer films with higher critical temperatures.\\
Besides the deterioration of the internal quality factors, we found a much smaller driving power bearable by the resonators before they bifurcate. Since the noise contribution due to the cold HEMT amplifier scales with the driving power as $P_{drive}^{-1/2}$, the faint amount of power made the X-ray pulses to be overwhelmed by the noise, so no pulse was observed. For this reason we started to study the dependence of the resonance profile on the driving power.

\subsection{Electron-phonon coupling}

\begin{figure}[htbp]
\begin{center}
\includegraphics[width=\linewidth, keepaspectratio]{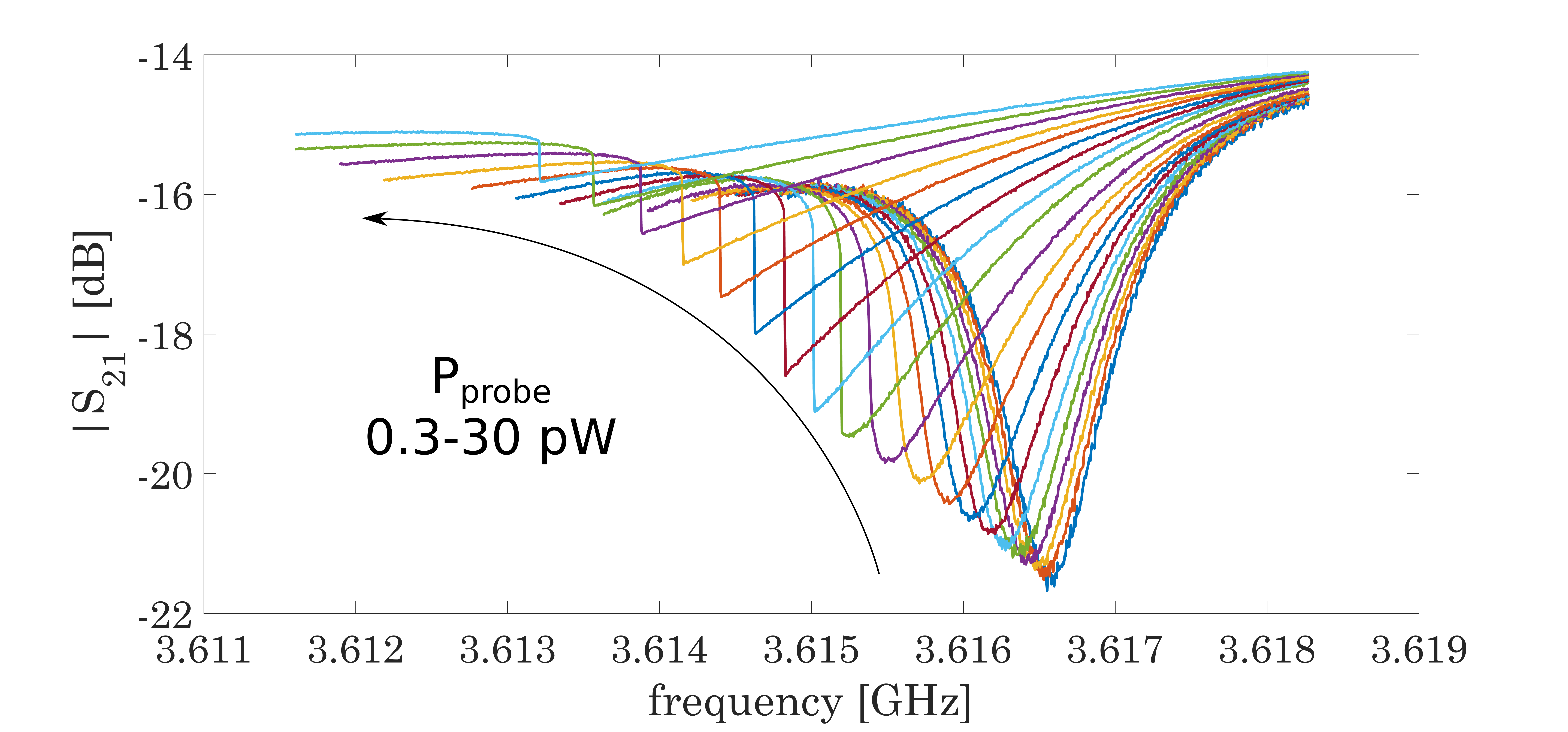}
\caption{Resonance profile acquired with a VNA at increasing readout powers: the compression of depth of the resonance is due to the worsening of the internal quality factor caused by the self-heating of the quasi-particles.}
\label{fig:fig3}
\end{center}
\end{figure}

A resonance can bifurcate mainly for two reasons: the intrinsic non linearity \cite{SWENSON} of kinetic inductance or because of self-heating of quasi-particles due to the dissipation of the readout power \cite{DEVISSER}. Both the effects result in a hysteretic behavior of the resonator. In our specific case, despite the energy associated to a single readout photon is at least one order of magnitude lower than the binding energy of the Cooper pairs, we observed a worsening of the internal quality factor while increasing the readout power (Fig.~\ref{fig:fig3}), which is compatible with a dissipative mechanism.

For this reason we decided to investigate the electron-phonon coupling in our Ti/TiN multilayer film. In order to do so, one should be able to measure the electrons temperature for each readout power (which in turn corresponds to a dissipated power within the resonator\cite{LEDUC}: $P_{diss}=2\frac{Q_r^2}{Q_cQ_i}P_{drive}$, where $Q_r$, $Q_c$ and $Q_i$ are the resonator, coupling and internal quality factors, respectively) and find in this way the coefficients of the power law\cite{IRWIN}:

\begin{equation}\label{eq:eqn1}
    P_{diss}=\Sigma V (T_{el}^n-T_{bath}^n)
\end{equation}
where $V$ is the active volume, $\Sigma$ is a material dependent coefficient, $T_{el}$ and $T_{bath}$ are the electrons and bath (i.e. phonons) temperatures, respectively; finally, the temperature exponent $n$ is determined by the heat transport mode and it is expected to be 5 when the thermal conductivity is dominated by electron-phonon conduction.

\begin{figure}[htbp]
\begin{center}
\includegraphics[width=\linewidth, keepaspectratio]{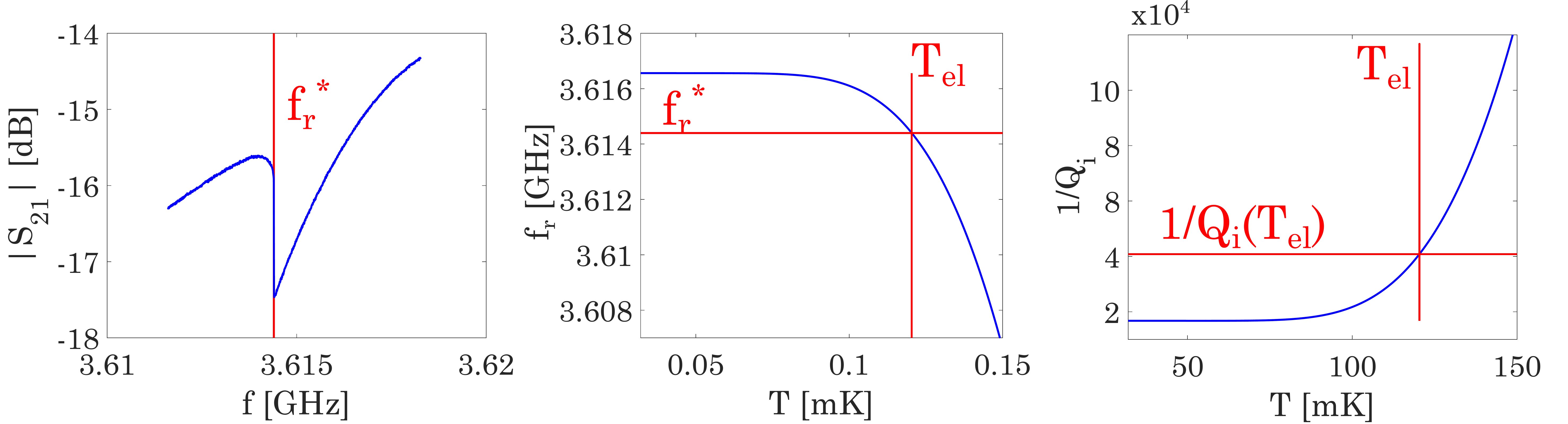}
\caption{Concept for obtaining the electrons temperature and internal quality factor to build the $T_{el}$ vs $P_{diss}$ plot: the electrons temperature is extrapolated from the calibration curve $f_r$ vs $T$ measured in the low power limit (where the approximation $T_{\text{bath}}\approx T_{\text{electrons}}$ is valid). Knowing $T_{el}$ it is possible to extrapolate the internal quality factor relative to that temperature.}
\label{fig:fig4}
\end{center}
\end{figure}

In order to extract the dissipated power and the electrons temperature, we adopt the procedure that is described in the following. The shape of the resonance passed the bifurcation is the result of the collection of points belonging to the same resonance which shifts under the influence of the probe signal. In particular, we assume that the frequency at which the resonance snaps ($f_r^*$ in the following) can be seen as the frequency where the ``real" resonance profile underneath has its own minimum.\\
The first step consists in performing a sweep of the bath temperature and acquire the resonance profile in the low power limit, in order to be sure to not affect the resonance shape: in this condition the electrons and bath temperatures do not differ significantly. This sweep is used to build the calibration curves of the resonant frequency and internal quality factors as a function of the electrons temperature. Given $f_r^*$ from the resonance profile, it is possible to extrapolate the electrons temperature ($T_{el}$) from the first curve (Fig.~\ref{fig:fig4}, {\it center}) while from the second one (Fig.~\ref{fig:fig4}, {\it right}) it is possible to extrapolate the internal quality factor ($Q_i(T_{el})$). Knowing $T_{el}$ and $Q_i(T_{el})$ for several $f_r^*$ (i.e. several readout powers) it is then possible to build the pairs of data to be fitted by the power law Eq. (\ref{eq:eqn1}).

\begin{figure}[htbp]
\begin{center}
\includegraphics[width=0.7\linewidth, keepaspectratio]{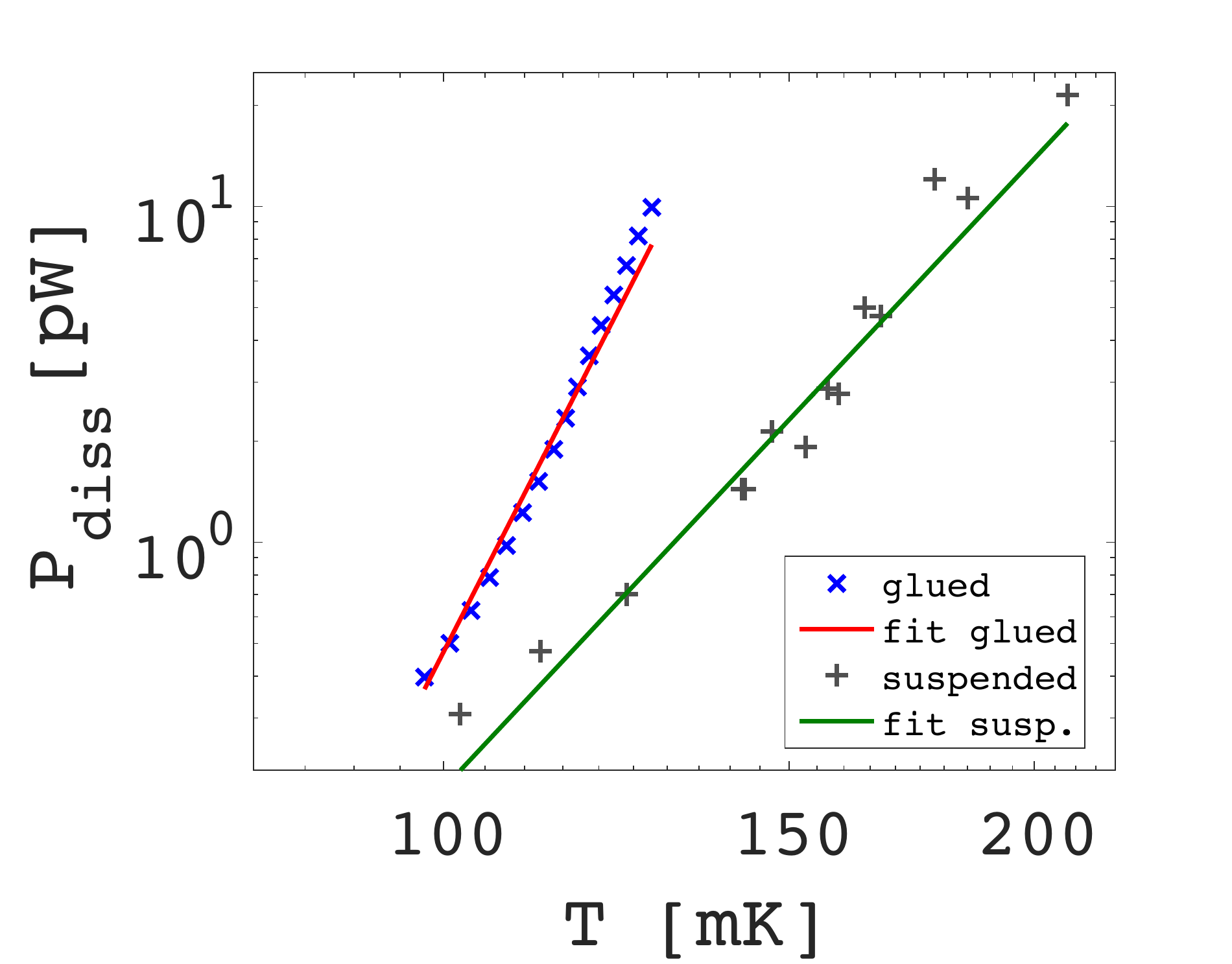}
\caption{Set of dissipated powers at different electrons temperatures extrapolated with the method described in the text obtained with the chip glued on the gold layer of the CPW (data: blue crosses - fit: red line) and with the chip suspended by the Al bonding wires (data: dark gray crosses - fit: green line). The deviation at high temperatures from the power law obtained by fitting the data (red line) is due to the fact that the functions used to fit the calibration data (Fig.~\ref{fig:fig4}, {\it right}) consist of approximations valid in the limit $k_BT\ll \Delta$. (Color figure online)}
\label{fig:fig5}
\end{center}
\end{figure}

In order to measure the electron phonon coupling, the bath temperature must be known precisely, so we modified the setup described in sec.~\ref{sec:sec2}: the chips in this scenario were glued directly onto the gold of the ground plane of the CPW, so that they shared the same temperature of the detectors holder. We acquired the resonance profile with a bath temperature of 32 mK for readout powers ranging between -46 and -31 dBm with steps of 1 dBm: the resulting plot is shown in fig.~\ref{fig:fig5}; the fit of the data resulted in a power coefficient  $n=11.4$, which is very different from the expected value of 5. The coefficient $\Sigma V$, instead, results to be 0.13 W/K$^n$.

As countercheck, this method to measure electron-phonon coupling has been tested with the original configuration described above in sec.~\ref{sec:sec2}. In this way, assuming that the thermal coupling provided by the two Al bonding wires is much weaker than the coupling between electrons and phonons in the Ti/TiN film, one should be able to determine the power coefficient expected for thermal conductance through metals: in this case (Fig.~\ref{fig:fig5}) $n$ turned out to be 6.1, which is still overestimated respect to what expected.
\\This systematic overestimation might be due to the fact that the law expressed by Eq. (\ref{eq:eqn1}) is valid while considering the number of quasiparticles responsible for thermal dissipation constant. In the case of superconducting metals, the population of quasiparticles is strongly dependent on the temperature through an exponential function, adding in this way a variable term that is not taken into account in this naive analysis.

\section{Conclusions}

In our attempt to use microwave microresonators as thermal detectors we found that resonators made of low critical temperature films of Ti/TiN multilayer feature lower quality factors than expected and at the same time the readout power bearable by these devices is not enough to detect signals coming from 6 keV X-rays.
\\We developed a method to measure the electron-phonon coupling which is capable to qualitatively discern the heat transport model, but for more quantitative conclusions a more complete and sophisticated model is required.

\begin{acknowledgements}
This work is supported by Fondazione Cariplo, through the project {\it Development of Microresonator Detectors for Neutrino Physics} (grant {\it International Recruitment Call 2010}, ref. 2010-2351).
\end{acknowledgements}

\pagebreak

\end{document}